\documentclass{article}
\usepackage{authblk}

\usepackage[margin=1in]{geometry}

\usepackage{libertine}
\usepackage[libertine]{newtxmath}
\usepackage[T1]{fontenc}

\usepackage{cite}
\usepackage{amsmath,amsfonts}
\usepackage{bm}
\usepackage{mathtools}
\usepackage{graphicx}
\usepackage{textcomp}
\usepackage{balance}
\usepackage[utf8]{inputenc}
\usepackage[ruled,vlined,linesnumbered,noresetcount]{algorithm2e}
\usepackage{hyperref}
\usepackage{xurl}
\usepackage{comment}
\usepackage{tabularx}
\usepackage{microtype}

\usepackage[dvipsnames]{xcolor}

\DeclarePairedDelimiter\abs{\lvert}{\rvert}%

\renewcommand{\Re}{\mathsf{Re}}

\def\BibTeX{{\rm B\kern-.05em{\sc i\kern-.025em b}\kern-.08em
    T\kern-.1667em\lower.7ex\hbox{E}\kern-.125emX}}
\usepackage{lipsum}

\title{Load Block Modeling in Distribution Systems: Network Reconfiguration for Load Restoration}

\author[1]{David M. Fobes}
\author[2]{Harsha Nagarajan}
\author[1]{Manuel Garcia}
\author[1]{Robert Ferrando}
\author[2]{Russell Bent}
\affil[1]{A-1 Computational Intelligence \& Modeling, Los Alamos National Laboratory}
\affil[2]{T-5 Applied Mathematics \& Plasma Physics, Los Alamos National Laboratory}
\date{\today}

\begin{document}

\maketitle

\tableofcontents

\newpage

\begin{abstract}
The distribution system restoration (DSR) problem has received considerable attention over the last decade or more. Solutions to the DSR problem identify the best set or sequence of actions to perform on a distribution circuit to restore service after a disruption. The problem is challenging from a computational perspective, with engineering constraints specific to distribution systems, such as radial operations, that are difficult to effectively model. In this paper, we revisit the model for how specific loads are shed, energized and restored--and develop a formulation that more accurately models the requirements of load shedding, load energizing and restoration in distribution systems.
\end{abstract}

\section{Introduction}

Resilience to extreme events and adverse conditions has been a driver for research and development in new ways to model, optimize, and control power systems. One crucial element of resilience is response to extreme events, and ultimately, restoration. During an event, or immediately after, operators of power systems determine sequences of actions to limit the impacts induced by the event.  While operators must respond and restore service at all levels of a power system, most events that lead to outages directly impact lower voltage distribution systems \cite{form417}. This observation has yielded extensive literature into developing approaches to solve what is referred to as the distribution restoration problem (DSR). The DSR considers distribution feeder(s) that have been degraded (e.g., physically damaged, faults, etc.), and, often, disconnected from the bulk electric transmission system. The goal of the DSR is to determine operator actions utilizing resources on the feeders (distributed generation, storage, network reconfiguration, microgrids, etc.) to energize and restore service to as much of the system as possible, until the degradation can be resolved (e.g., repairs, fault clearing, etc.).

This paper considers two specific, critical pieces of the DSR problem definition: 1) the modeling of loads and how loads are shed, energized, and restored and 2) the modeling of grid-forming distribution generator (DG) requirements when isolated from the bulk electric system (BES). {Considering the distribution system in restoration, particularly with the potential for favorable interaction with networked microgrids, provides an additional layer of granularity in operational decisions that may achieve global objectives, including a more practical, realistic means of load shedding \cite{schneider2021t}, \cite{bent2022integrated}.} First, when considering the literature on DSR, many studies model loads as individually controllable, which implicitly assumes meter-based controls, such as advanced metering infrastructure (AMI), to determine which loads receive power and which do not. While most AMIs are limited to coarse degrees of freedom (allow power to flow to the load or not), often, this model of loads is further simplified to allow continuous load control. In practice, assumptions that loads are controllable individually and may be controlled through a demand response-like technology are often unrealistic.
In many feeders, deployed AMI technologies are limited and a load is energized in its entirety if connected to an operating energy resource through a closed circuit. 
Despite these limitations, continuous load control has been an attractive approximation, as it reduces the computational challenges associated with solving the DSR. Most notably, such approximations eliminate the need for modeling load connectivity to energy resources and eliminates the binary variables that are used to determine whether a load is shed or not. 

Second, within the DSR literature, many studies have also shown the value of using DGs to improve restoration, esp. when islanded from the BES \cite{xu2017dgs,kim2016framework,sharma2015decentralized}. However, most of these studies assume that any DG  is sufficient to re-energize a disconnected load. In practice, it is necessary that the load also be connected to one (and only one) DG  with an active grid-forming inverter with sufficient capabilities to form an islanded grid. Like load shedding, this detail has been neglected as it introduces additional computational and modeling complexity. 

To address these challenges, this paper develops an optimization formulation 
that is computationally effective at modeling these core requirements of load and   DG  control, i.e., load cannot be shed unless it is completely disconnected from sources of power through network reconfiguration and it can only be energized if it connected to a   DG  with grid-forming capabilities.  This formulation dramatically reduces the number of binary variables required to model on-off controls of individual loads by clustering loads into blocks that are always energized in tandem, as defined by network topology, available controls, and access to grid-forming technologies.  In short, the key contributions of the paper include:

\begin{itemize}

\item A formal mathematical model for load shed based on connectivity to a DG. The model strictly reduces the number of binary variables used to turn on or off loads by grouping loads together into topologically defined \emph{load blocks}.

\item A formal mathematical model for grid-forming inverters which only allows loads to be energized if they are topologically connected to a   DG  with grid-forming capabilities.

\item A fundamental generalization of individual load control as meter-based, on-off control can be crudely thought of as a network configuration option. While such meter-based, individual load control does not benefit from binary variable reduction, this formulation is a useful construct for modeling the individual load control scenario.

\item Empirical results obtained using PowerModelsONM.jl \cite{9897093} which demonstrate a notable improvement in solution quality and accuracy over the more traditional load approximation methods, without significant increases (and even decreases) in computational run times.

\end{itemize}

The rest of this paper is organized as follows. Section \ref{sec:lit} provides a literature review of DSR in the context of load modeling. Section \ref{sec:formulation} covers the details of the DSR formulation. Sections \ref{sec:load} and \ref{sec:grid-forming} describe the new formulations for load shedding and grid-forming inverters, respectively. Section \ref{sec:results} provides experimental results to corroborate the efficacy of the approach and Section \ref{sec:conclusion} concludes the paper.

\section{Literature Review}\label{sec:lit}

The literature on the DSR (and related problems under different names) spans at least the past decade. The literature can be categorized in a variety of ways, including those that focus primarily on DGs \cite{sharma2015decentralized}, \cite{xu2017dgs}, \cite{hafez2016decentralized}, \cite{sharma2016decentralized}, microgrids \cite{7513408}, \cite{li2014distribution}, \cite{chen2015resilient}, \cite{7971935}, and networked microgrids \cite{8606281}, \cite{kim2016framework}, \cite{wang2015self}, \cite{arif2017networked}, \cite{gao2016resilience}. However, for the purposes of this paper, the most relevant characteristic is how loads are modeled for shedding, energizing, or restoration in DSR.

The computationally easiest way to model load shedding and restoration is to introduce continuous variables for loads that allow the power consumed by a load to vary between 0 and its desired amount. The literature includes quite a few papers that model load in this way, including, but not limited to, \cite{Ferreira2014,Xu2001,Mogaka2020,lopes2014operation,sharma2015decentralized,wang2015self,arif2017networked,9477426}. While computationally advantageous, continuous loads are unrealistic in practice unless the distribution feeder has a demand response-like technology that allows for direct control of a load's power consumption.  However, it is important to note that this model is formally a relaxation of the more complex mathematical formulations discussed later in this section's review. Thus, it is an upper bound for restoration that is achievable in practice.  

To address the practicality of modeling loads as continuously shedable components, some literature has turned to using binary variables to model how much power a load consumes. In other words, a load may consume nothing or consume its desired amount, nothing in between.  Examples of papers that model loads in this way include (not exhaustively) \cite{8606281,xu2017dgs,sharma2016decentralized,8606281,chen2015resilient}. Due to the combinatorics introduced with binary variables, modeling binary loads is generally more computationally challenging.  While solutions based on this model are more practical than the continuous model, it still assumes that there is AMI or controllable switches at each load that allow loads to be individually disconnected from the distribution feeder. Once again, this load control model is a relaxation of the more complex formulations discussed next.

Before turning to mathematical formulations that model a load's consumption status explicitly by whether or not it is connected to an energy source, we first discuss those papers that model load connections implicitly through the design of algorithms that were developed to solve the DSR.  Examples include \cite{hafez2016decentralized,7513408,li2014distribution,gao2016resilience} and a common feature of these papers are heuristics that sequentially restore loads by reconnecting them in sequence starting from an energy resource (typically, the connection to the bulk system).  The exception is \cite{li2014distribution}, which enumerates the possible network configurations and loads are implicitly energized depending on the selected configuration.  While a notable advance in the state-of-practice, these heuristics have no guarantees on achieving the globally optimal solution to the DSR. An explicit mathematical formulation, such as described by this paper, supports advances in finding globally optimal solutions to DSR that model the connections between loads and resources.

The most closely related work to this paper is the sequence of papers \cite{7971935,Lei2020,7971945,8587147}. To the best of our knowledge, this set of papers, by the same sets of authors, are the first to introduce mathematical formulations that tie a load's status to whether or not it is connected to an energy resource that is active. The specifics of the formulation vary (for example, \cite{8587147} enumerates paths between loads and energy resources), but a common feature is the introduction of binary variables for all components (loads, buses, etc.) to represent their energization state. In contrast, we reduce the number of required binary variables 
to improve the computational complexity of these models. In addition, none of these papers model requirements on including a grid-forming inverter in any energized portion of the feeder.

In addition, we highlight the work of \cite{Lei2020}, which, like our paper, revisits a piece of the DSR that is challenging to model from a computation standpoint.  The work of \cite{Lei2020} complements this paper, as it focuses
on developing effective mathematical formulations (constraints) to enforce radial topologies when the network is re-configured. Our inspiration is to focus on computationally tractable formulations for modeling load control, which is similar to these motivations of \cite{Lei2020}.  

{  Finally, the load shedding and grid-forming inverter models introduced in the present paper have been implemented and employed in \cite{Moring_2024} and \cite{Moring_2025}, which both consider the operation of networked microgrids under uncertainty. The former article \cite{Moring_2024} formulates a robust optimization problem, while the latter \cite{Moring_2025} enhances the practicality of the former's robust optimization implementation and performs OPF to interpolate between topology changes. However, both of these articles are primarily focused on the empirical utility of the models. By contrast, this paper presents a theoretical framework for the grid-forming inverter model, and demonstrates the superiority of this model, combined with load shedding in blocks, in increasing overall load satisfaction.}

\section{Distribution System Restoration Formulation}\label{sec:formulation}

In this section, we discuss the basic DSR problem formulation, which improves the fidelity and practicality of the DSR problem and highlights the utility of the forthcoming load shedding model. 

\subsection{Three Phase Power Flow}
\label{sec:three-phase-power-flow}

In unbalanced, 3-phase modeling, the power flow (Ohm's Law) on a line $(i,j,k)$ and phase, $\phi$, is expressed with

\begin{equation}
     {S}^{\phi}_{k} =  V^{\phi}_i\sum_{\hat{\phi} \in \Phi_{k}}\left(\bm{Y}^{\phi\hat{\phi}}_{k}\right)^*\left(V^{\hat{\phi}}_i - V^{\hat{\phi}}_j\right)^*,  \forall (i,j,k) \in \cal E, \label{acopf:ohms}
\end{equation}

\noindent
 where $(\cdot)^*$ denotes conjugate transpose.  Here, $\cal E$ denotes the set of all edges (power lines), as defined by $(i,j,k)$--a unique identifier, $k$, and the two buses the line is connected to, $i$ and $j$. Then, $S$ denotes the complex power flow on a line and $\bm{Y}$ denotes the complex admittance of a line, as indexed by $k$, and $V$ denotes the complex voltage at a bus, as indexed by $i$ and $j$.
 $\Phi = \{a,b,c\}$ is used to denote the set of phases and each quantity is also indexed by phase(s), $\phi \in \Phi$. Flow balance at each bus (Kirchhoff's Law) is expressed with 

\begin{equation}
    \sum\limits_{d \in {\cal D}(i)}{S}^{\phi}_d - {{S}}^{\phi}_i =
    \sum\limits_{(i,j,k) \in {\cal E}_i^+} S^{\phi}_{k} - \sum\limits_{(j,i,k) \in {\cal E}_i^-} S^{\phi}_{k},  \forall i \in {\cal N}, \phi \in \Phi, \label{acopf:kirchhoff}
\end{equation}

\noindent Here, ${\cal D}(i)$ denotes the set of   DG  s at bus $i$, $S_d^\phi$ denotes the power produced by   DG  $d$, and $S_i^\phi$ the power consumed at bus $i$. The sets ${\cal E}_i^+$ and ${\cal E}_i^-$ refer to the edges connected to bus $i$ as oriented ``from $i$'' and ``to $i$'', respectively.

\subsection{Temporal Power Flows}
The DSR models power flows over time as a sequence of states that reflects changes in the system, such as network reconfiguration. Given a set of time points, $\cal T$, indexed by $t$, equations \eqref{acopf:ohms} and \eqref{acopf:kirchhoff} are restated as

\begin{subequations}\label{acopf:power_time}
\begin{equation}
     {S}^{\phi}_{kt} =  V^{\phi}_{it}\sum_{\hat{\phi} \in \Phi_{k}}\left(\bm{Y}^{\phi\hat{\phi}}_{k}\right)^*\left(V^{\hat{\phi}}_{it} - V^{\hat{\phi}}_{jt}\right)^*, \label{acopf:ohms_time}
\end{equation}
\begin{equation}
    \sum\limits_{d \in {\cal D}(i)}{S}^{\Phi}_{dt} - {{S}}^{\Phi}_{it} =
    \sum\limits_{(i,j,k) \in {\cal E}_i^+} S^{\Phi}_{kt} - \sum\limits_{(j,i,k) \in {\cal E}_i^-} S^{\Phi}_{kt}, \label{acopf:kirchhoff_time}
\end{equation}
\end{subequations}

\noindent $\forall t \in \cal T$. To simplify notation, the temporal index $t$ is omitted from the subsequent constraint definitions, except where constraints link time periods.  

\subsection{Engineering Limits}

Each of the quantities described in equations \eqref{acopf:power_time} are bounded by engineering based limits.
These include bus voltage magnitude constraints~\eqref{acopf:volt},   DG  constraints~\eqref{acopf:derlimit}, load consumption constraints~\eqref{acopf:demlimit}, thermal limit constraints~\eqref{acopf:thermal} and phase angle difference constraints~\eqref{acopf:pad}:

\begin{subequations}\label{acopf-engineering}
\begin{equation}
  \underline{\bm  v}_{i} \le \abs{V_{i}}^\phi \le \overline{\bm  v}_{i}, \;\;\;\; \forall ~i \in {\cal N},~\phi \in \Phi_d\label{acopf:volt}
 \end{equation}
 \begin{equation}
  \underline{\bm  S}_d  \le |S_{d}|^\phi \le   \overline{\bm S}_d, \;\;\; \forall ~d \in {\cal D},~\phi \in \Phi_i\label{acopf:derlimit}
 \end{equation}
 \begin{equation}
 0  \le |S_{i}|^\phi \le \overline{\bm S}_{i}, \;\;\; \forall ~i \in {\cal N},~\phi \in \Phi_i \label{acopf:demlimit}
 \end{equation}
\begin{equation}
 \abs{S_{k}}^\phi \le  \overline{\bm S}_{k}, \;\;\; \forall (i,j,k) \in {\cal E},~\phi \in \Phi_{k}  \label{acopf:thermal}
 \end{equation}
\begin{equation}
 \underline{\bm  \theta}_{k}  \le \angle(V_{i}^\phi {V_{j}^\phi}^*) \le  \overline{\bm \theta}_{k}, \;\;\; \forall (i,j,k) \in {\cal E},~\phi \in \Phi_{k} \label{acopf:pad},
 \end{equation}
\end{subequations}

\noindent 
Here, the notation $\underline{(\cdot)}$ and $\overline{(\cdot)}$ is used to denote the lower and upper engineering limit on a quantity and $\angle(\cdot)$ is used to denote the angle that a complex number has in the complex plane.

\subsection{Network Configuration}

Distribution feeders include switches and reclosers for opening and closing circuits. They are used to restore loads, back-feed portions of a network, and isolate faults. Given a set of power lines that are capable of opening and closing, $\Gamma \in \cal E$, the variable $\gamma \in \{0,1\}$ is used to denote whether or not the line is open ($\gamma = 0$) or closed ($\gamma = 1$). The equations for power across these edges (Equations \eqref{acopf:ohms_time}) are then restated as

\begin{equation}
     {S}^{\phi}_{k}  = \gamma_{k} V^{\phi}_{i}\sum_{\hat{\phi} \in \Phi_{k}}\left(\bm{Y}^{\phi\hat{\phi}}_{k}\right)^*\left(V^{\hat{\phi}}_{i} - V^{\hat{\phi}}_{j}\right)^*,  \label{acopf:power_sum3_time_on_off} 
\end{equation}

\noindent 
$\forall (i,j,k) \in {\Gamma}, \phi \in \Phi_{k}$.
Similarly, Equation~\eqref{acopf:pad} is modified to deactivate the phase angle difference constraint when the edge is opened,

\begin{equation}
\underline{\bm \theta} (1-\gamma_{k})  +  \underline{\bm  \theta}_{k}  \le \angle(V_{i}^\phi {V_{j}^\phi}^*) \le  \overline{\bm \theta}_{k} + \overline{\bm \theta} (1-\gamma_{k}) ~\label{acopf:pad_on_off},
\end{equation}

\noindent $\forall (i,j,k) \in {\Gamma},~\phi \in \Phi_{k}$, where $\underline{\bm \theta}$ and $\overline{\bm \theta}$ are the minimum and maximum possible voltage angle difference between any two buses in the system.

\subsection{Battery Storage}

Equations \eqref{constraint:bat1}-\eqref{constraint:bat3} model the operating limits of batteries and their storage capacity. The set of batteries is denoted by $ {\cal B} \in {\cal D}$.
Constraint \eqref{constraint:bat1} computes the stored energy ($\psi$) at each time $t$, where $\tilde{p}$ is the rate at which the energy is extracted (stored) and $\Delta t$ is the duration of time that this extraction/storage rate occurs (the length of a time step).  Constraint \eqref{constraint:bat2} ensures that the stored energy never exceeds the limits of the battery. Constraint~\eqref{constraint:bat3} computes the power output based on piece-wise energy loss function, $l \in L$ \cite{Geth2020}.
\begin{subequations}\label{constraint:battery}
\begin{equation}
\psi_{\beta t}  = \psi_{\beta t-1} - \tilde{p}_{\beta t} \Delta t,  \;\;\;\forall \, t \in {\cal T},  \beta \in {\cal B}, \label{constraint:bat1} 
\end{equation}
\begin{equation}
0  \leq \psi_{\beta t} \leq \overline{\bm \psi}_{\beta},  \;\;\;\forall \, t \in {\cal T},  \beta \in {\cal B}, \label{constraint:bat2} \\
\end{equation}
\begin{equation}
p_{\beta t}  \le l_{\beta}^{} \tilde{p}_{\beta t} + l_{\beta}^{'}, \;\;\;\forall \, t \in {\cal T},  \beta \in {\cal B}, l \in L_\beta. \label{constraint:bat3}
\end{equation}
\end{subequations}
\subsection{Radial Operations}
Distribution systems must typically operate with a radial structure, which is enforced with the following constraints:
\begin{subequations}\label{constraint:radial}
\begin{flalign}
& \mathbf{\omega} \in \mathbf{\Omega}, \;\;\;  \forall(i,j,k) \in \Gamma \label{eq:radial1}, \\
& \gamma_{k} \leq \omega_{k},  \;\;\;  \forall(i,j,k) \in \Gamma. \label{eq:radial2}
\end{flalign}
\end{subequations}
\noindent 
In (\ref{eq:radial1})-(\ref{eq:radial2}), $\omega_{k}$ is an auxiliary variable for a \emph{virtual} state of switch $(i,j,k)$, and $\mathbf{\Omega}$ is a set of all incidence vectors of possible spanning tree (forest) topologies the network can form. Therefore, (\ref{eq:radial1}) enforces $\mathbf{\omega}$ to form a spanning tree, and (\ref{eq:radial2}) enforces $\gamma$ to form a sub-graph of a possible spanning tree. Reference \cite{Lei2020} describes and compares different ways of implementing this constraint.

\subsection{Objective}
The overall objective of the DSR is to maximize load that is energized, as stated by
\begin{equation}
\sum_{t \in \cal T} \sum_{i \in {\cal N}} \sum_{\phi \in \Phi_i} \kappa_i \,\Re\left(S_{it}^\phi\right)  \label{obj},
\end{equation}

\noindent where $\kappa$ is a load-specific weighting term that allows the prioritization of energizing a particular load. The basic DSR problem is then stated as

\begin{center}
\begin{tabular}{rl}
$\mathrm{max}$ & \eqref{obj} \\
$\mathrm{s.t.}$ & \eqref{acopf:kirchhoff_time}, \eqref{acopf-engineering}, \eqref{acopf:power_sum3_time_on_off}, \eqref{acopf:pad_on_off}, \eqref{constraint:battery}, \eqref{constraint:radial}.
\end{tabular}
\end{center}

\section{Load Shedding Model}\label{sec:load}

In section \ref{sec:formulation}, the basic DSR allows continuous load shedding as defined by constraints \eqref{acopf:demlimit}. However, without advanced demand response-like technologies, it is typically unrealistic to expect that load can be shed at this level of granularity in distribution circuits. Thus, many DSR strategies introduce binary variables $z \in \{0,1\}$ to control whether a load is fully energized or not and change constraints \eqref{acopf:demlimit} to

\begin{equation}\label{acopf:demlimit_onoff}
     S_{i}^\phi = z_{i} {\bm S_{i}}^\phi. 
\end{equation}

\noindent In practice, such a model assumes that advanced metering is available or that there are remotely controllable switches at every load. When such components are unavailable, load is only shed when there is not a closed circuit to a DG and the load cannot be shed if connected to an active distributed generator. Thus, it is necessary to model the connectivity between loads and distributed generators. To accomplish this, we introduce one of the the primary contributions of this work--\emph{a new model for load shedding based on a set $\cal Z$}. Each element of $\zeta \in {\cal Z}$ is a set of components, referred to as a load block, that are connected together when all switches in the network are open. Here, a load block is purely defined by the topological structure of the network, however, without loss of generality arbitrary load blocks, for example, as defined by an operator, could be introduced by redefining $\cal Z$. 

By definition, if one load in the load block is energized, then all loads in the block are energized.  Given $\cal Z$, the determination of whether or not a load is energized is modeled with constraints:
\begin{subequations}\label{constraint:load_block}
\begin{equation}\label{constraint:load_block_der}
    z_{\zeta} \ge \frac{S_{d}^\phi}{\overline{\bm S_d}} 
\end{equation}
\begin{equation}\label{constraint:load_block_neighbor}
     z_{\hat{\zeta}} - (1-\gamma_{k}) \le z_{\zeta} \le  z_{\hat{\zeta}} - (1-\gamma_{k})
\end{equation}
\begin{equation}\label{acopf:load_block_onoff}
     S_{i}^\phi = z_{\zeta} {\bm S_{i}}^\phi. 
\end{equation}
\end{subequations}
\noindent where $z$ is redefined as a binary variable to indicate whether or not a block, $\zeta$, is energized.  Here, constraint \eqref{constraint:load_block_der} is defined $\forall \zeta \in {\cal Z}$, $\forall \phi \in \Phi$, and $\forall d \in \zeta$ (the distributed generators in load block $\zeta$). This constraint forces the load block to be energized if any   DG  in the block is producing power. Constraint \eqref{constraint:load_block_neighbor} is defined $\forall \zeta, \hat{\zeta} \in {\cal Z}$, where  $\zeta$ and $\hat{\zeta}$ are connected by a switch $(i j k)$. 
This constraint forces neighboring blocks to have the same energization state when connected by a closed switch. Thus, if a   DG  is producing power, constraint \eqref{constraint:load_block_der} will energize the DG's block, constraint \eqref{constraint:load_block_neighbor} will energize all the adjacent blocks (as determined by the switch state), constraint \eqref{constraint:load_block_neighbor} will further energize the neighbors of the adjacent blocks, and so forth, implicitly modeling the paths between DGs and each block. Constraint \eqref{acopf:load_block_onoff} then restates constraint \eqref{acopf:demlimit_onoff} to use the load block variable $z_\zeta$ for all buses $i \in \zeta$.

\section{Grid-forming Inverter Model}\label{sec:grid-forming}

While the load shedding model introduces topological decisions to determine whether or not a load is energized, it does not model whether or not connected load blocks are supported by a grid-forming inverter. In this section, we present the second key contribution of this work--\emph{a new model for enforcing that every connected set of load blocks has exactly one grid-forming DG}\footnote{Note that the model can be relaxed to be ``at least one'' grid-forming DG, where, without loss of generality, post-processing can select which DG  operates in grid-forming mode. Limiting to one grid-forming inverter reduces the search space by eliminating redundant equivalent solutions.}.  This model is inspired as a variant of graph coloring formulations \cite{jensen2011graph} where connected nodes must have the same color rather than different colors.

First, for every switch $ijk \in \Gamma$, we assign a variable ``color'' $y^\zeta_{k}$, where the color $\zeta$ is used to model which load block ($\zeta$) with a grid-forming inverter is topologically connected to $k$. More formally,

\begin{equation} \label{constraint:SwitchColor}
    \sum_{\zeta \in {\cal Z}} y^\zeta_{k} \le \gamma_{k}
\end{equation}

\noindent states every switch can have one color when open and has no color when closed. Next, for each load block, $\zeta \in {\cal Z}$ with at least one   DG  with a grid-forming inverter, the requirement for having at most one grid-forming inverter active is modeled with the constraint,

\begin{equation} \label{constraint:DER}
    \sum_{ijk \in {\Gamma}_\zeta} (1-\gamma_{k}) - |{\Gamma}_\zeta| + 1 \le \sum_{d \in \zeta} x_d \le 1,
\end{equation}

\noindent  where $\Gamma_\zeta$ defines the switches connected to $\zeta$ and $x$ is a binary variable associated with every   DG  to determine whether it is being used in grid-forming mode ($x=0$ for any   DG  without grid-forming capabilities). The right-hand side of this constraint ensures only one DG is grid-forming.  The left-hand side ensures that one DG is grid-forming if $\zeta$ is isolated from the the rest of the grid. 
Now, for each load block $\zeta \in \cal Z$, and $ijk \in \Gamma_\zeta$, the constraint

\begin{equation} \label{constraint:NeighborColor}
    y^\zeta_{k} - (1 - \gamma_{k}) \le \sum_{d \in \zeta} x_d \le  y^\zeta_{k} + (1 - \gamma_{k}),
\end{equation}

\noindent
is used to color all closed switches connected to $\zeta$ with $\zeta$ when $\zeta$ has an active grid-forming DG. This constraint eliminates $\zeta$ as a color otherwise. 
Then, for every load block $\zeta \in \cal Z$, $(ijk), (\hat{i}\hat{j}\hat{k}) \in \Gamma_\zeta$, such that $k \ne \hat{k}$, the constraint

\begin{subequations} \label{constraint:CoupleColor}
\begin{equation} 
  y^{\hat{\zeta}}_{\hat{k}} - (1 - \gamma_{\hat{k}}) - (1 - \gamma_{k}) \le y^{\hat{\zeta}}_{k},
\end{equation}
\begin{equation} 
 y^{\hat{\zeta}}_{k} \le  y^{\hat{\zeta}}_{\hat{k}} + (1 - \gamma_{\hat{k}}) + (1 - \gamma_{k}),
\end{equation}
\end{subequations}

\noindent
$\forall \hat{\zeta}$, is added to ensure that all closed switches connected to $\zeta$ have the same color.  Together, constraints \eqref{constraint:SwitchColor}-\eqref{constraint:CoupleColor} have a propagating effect that forces all switches of connected load blocks to have the same color. 

The next set of constraints enforce that this color must be from one of the connected load blocks and uses a multi-commodity network flow model. First, constraint

\begin{equation}\label{constraint:LinkInvertorToColor}
y_{k}^\zeta \le \sum_{d\in \zeta} x_d, \;\;\; \forall \zeta \in {\cal Z}, ijk \in \Gamma_\zeta, 
\end{equation}

\noindent states that a switch connected to $\zeta$ cannot take color $\zeta$ unless one of the DGs of $\zeta$ is grid-forming.  
Next, let $\Upsilon
_\zeta$ be a set of virtual edges connecting $\zeta$ to all other load blocks $\hat{\zeta}$.  Then, for each switch $ijk \in \Gamma$, let $f_{k}^\zeta$ be a variable flow of color $\zeta$ on $k$ where

\begin{equation}\label{constraint:switchflow}
-\gamma_{k} |{\cal Z}| \le f_{k}^\zeta \le \gamma_{k} |{\cal Z}|.
\end{equation}

\noindent
Then, for $ab$, where $ab$ is a virtual edge between load blocks $a$ and $b$, let $\upsilon_{ab}^\zeta$ be a variable flow of color $\zeta$ on $ab$ where

\begin{equation} \label{constraint:virtual}
0 \le \upsilon_{ab}^\zeta \le 1.
\end{equation}

\noindent
These flow variables are used to determine if there is a connection between a grid-forming   DG  and a load block using the following constraints for each load block $\zeta \in {\cal Z}$,

\begin{equation} \label{constraint:source}
\sum_{ijk \in {\Gamma}_\zeta : i \in \zeta} f_{k}^\zeta - \sum_{ijkt \in {\Gamma}_\zeta : j \in \zeta} f_{k}^\zeta + \sum_{ab \in \Upsilon^\zeta} \upsilon_{ab}^\zeta = |{\cal Z}| - 1
\end{equation}

\begin{equation} \label{constraint:balance}
\sum_{ijk \in {\Gamma}_{\hat{\zeta}} : i \in \hat{\zeta}} f_{k}^\zeta - \sum_{ijk \in {\Gamma}_{\hat{\zeta}} : j \in \hat{\zeta}} f_{k}^\zeta  -  \upsilon_{\zeta \hat{\zeta}}^\zeta = -1, \;\;\; \forall \hat{\zeta} \ne \zeta
\end{equation}

\noindent
which state that load block $\zeta$ must send out enough flow such that every other load block gets 1 unit of flow (constraints \eqref{constraint:source}) and every load block consumes one unit of flow (constraints \eqref{constraint:balance}).  Then, for each $\zeta$, each $\hat{\zeta} \ne \zeta$ and each $ijk \in {\Gamma}_{\hat{\zeta}}$, constraint

\begin{equation}\label{constraint:flowlink}
y_{k}^\zeta \le 1 - \upsilon_{\zeta \hat{\zeta}}^\zeta
\end{equation}

\noindent
is used to force color $y_{ijkt}^\zeta$ to be unused for load block $\hat{\zeta}$ when there is flow on the virtual edge from $\zeta$ to $\hat{\zeta}$.  When $\zeta$ and $\hat{\zeta}$ are unconnected, the virtual edge is the only way to provide flow from $\zeta$ to $\hat{\zeta}$, so $\upsilon$ forces the corresponding $y$ to 0, thereby only allowing a connected set of load blocks to use a color within this set. In other words, activating a grid-forming inverter within these load blocks. 

The remaining sets of constraints tie   DG  usage and loads to these constraints. First,
for any DG, $d$, in load block $\zeta$,  the following constraints are added

\begin{equation}
S_{d}^\phi \le \overline{\bm S}_d \left(\sum_{ijk \in {\Gamma}_\zeta} {\gamma_{k}} + \sum_{j \in {\cal D}_\zeta} x_j\right) \label{constraint:grid_follow1}
\end{equation}

\begin{equation}
 S_{d}^\phi \le \overline{\bm S}_d \left(\sum_{ijk \in {\Gamma}_\zeta} \sum_{\hat{\zeta} \in {\cal Z}} y_{k}^{\hat{\zeta}} + { \sum_{j \in {\cal D}_\zeta} x_j}\right). \label{constraint:grid_follow4}
\end{equation}

\noindent Here, equation \eqref{constraint:grid_follow1} states that when all switches of a load block are open and all DGs in the load block are grid following, then there is no power output.  Constraint \eqref{constraint:grid_follow4} similarly restricts power output when no switch is assigned a grid-forming inverter from a connected load block.

\noindent The final DSR problem is then stated as

\begin{center}
\begin{tabular}{rl}
$\mathrm{max}$ & \eqref{obj} \\
$\mathrm{s.t.}$ & \eqref{acopf:kirchhoff_time}, \eqref{acopf:volt}, \eqref{acopf:derlimit}, \eqref{acopf:thermal}, \eqref{acopf:pad}, \eqref{acopf:power_sum3_time_on_off}-\eqref{constraint:grid_follow4}.
\end{tabular}
\end{center}

\section{Empirical Results}\label{sec:results}
\begin{center}
\begin{table}
\centering
\caption{\label{results-table} Empirical results for the three use cases and three formulations. ($P_d$ = load delivery)}
\bgroup
\def\arraystretch{1.5}
\setlength\tabcolsep{0.1cm}
\begin{tabular}{ p{1.8cm}<{\raggedright} p{0.7cm}<{\raggedleft}  p{0.9cm}<{\raggedleft}  p{0.65cm}<{\raggedleft} p{0.75cm}<{\raggedleft}  p{0.5cm}<{\raggedleft}  p{0.65cm}<{\raggedleft}  p{0.4cm}<{\raggedleft} p{0.75cm}<{\raggedleft} }
    \hline\hline
     case & bin. vars. & cont. vars. & solve time (s) & obj. & opt. gap (\%) & \# load shed & \# bl. shed & $P_{d}$ (MWh) \\
     \hline\hline
        ieee13-mod \mbox{traditional} & 818 & 9270 & 45.9 & 80.2 & 0.0 & 87 & 19 & 6.7 \\
        ieee13-mod \mbox{block} & 370 & 9198 & 3.48 & 253.0 & 0.0 & 136 & 35 & 1.9 \\
        ieee13-mod \mbox{block+gfm} & 786 & 10766 & 2.64 & 270.0 & 0.0 & 141 & 38 & 0.9 \\
        \hline
        iowa240-mod \mbox{traditional} & 16644 & 124284 & 59.0 & 49.1 & 0.0 & 163 & 64 & 44.1 \\
        iowa240-mod \mbox{block} & 1188 & 124068 & 2.64 & 516.0 & 0.0 & 2733 & 67 & 26.2 \\
        iowa240-mod \mbox{block+gfm} & 2772 & 129108 & 1.87 & 524.0 & 0.0 & 2781 & 68 & 25.5 \\
        \hline
        utility-feeder \mbox{traditional} & 81190 & 749634 & 3002 & 893.0 & 11.8 & 8392 & 455 & 17.2 \\
        utility-feeder \mbox{block} & 8998 & 748770 & 235.0 & 1100.0 & 0.0 & 15565 & 414 & 17.2 \\
        utility-feeder \mbox{block+gfm} & 63046 & 1310562 & 520.0 & 1100.0 & 0.0 & 15565 & 366 & 17.2 \\
        \hline\hline
\end{tabular}
\egroup
\end{table}
\end{center}
To demonstrate the efficacy of our approach to modeling loads we utilize three use cases, denoted henceforth as \emph{ieee13-mod}, \emph{iowa240-mod}, and \emph{utility-feeder}. Details of the three cases and the implementation are presented in the appendix. 

 \subsection{Limitations of Traditional DSR Models}

First, we consider the results of a traditional DSR optimization, which highlight two of the key limitations of this model 1) computational inefficiencies because of larger number of binary variables and 2) impractical load control decisions. As we can see from Table~\ref{results-table}, the number of loads and whole blocks shed over the restoration period is notably smaller than for the other DSR models, but the number of binary variables is considerably larger and the solve time is significantly longer in all cases. 
In particular, for the \emph{utility-feeder traditional} case the solver was only able to achieve an optimality gap of 11.8\% as of the $3000$s time limit, whereas the other two models achieved optimality in 520 seconds or less.

\begin{figure}
    \centering
    \includegraphics[width=0.7\columnwidth]{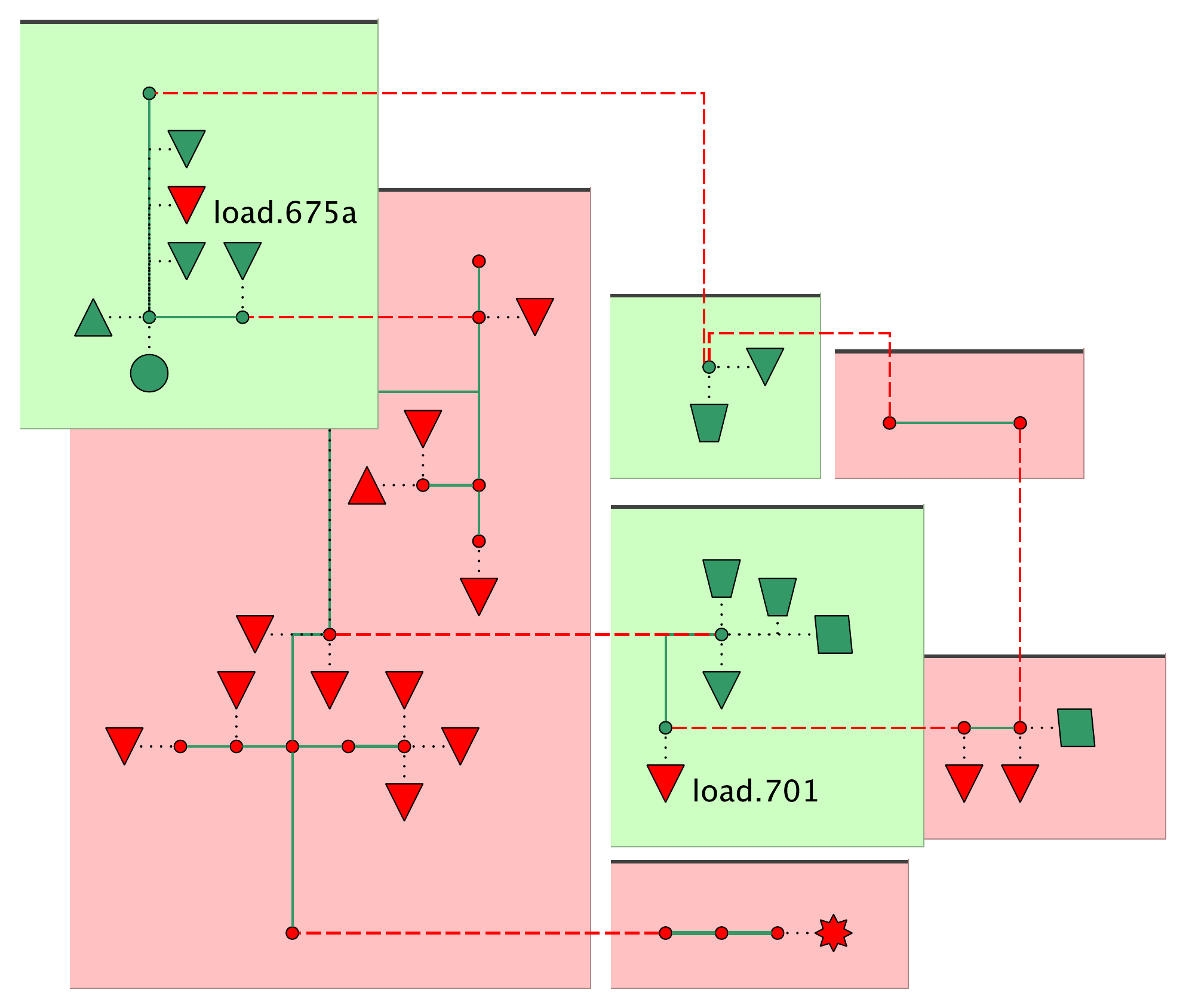}
    \caption{One-line diagram of the first time step of \emph{ieee13-mod traditional} case, where red objects/zones are shed objects/blocks, and green objects/zones are energized objects/blocks.}
    \label{fig-traditional}
\end{figure}

In Fig.~\ref{fig-traditional}, we show a case where the second limitation is demonstrated--individual loads have been shed inside energized blocks by the \emph{ieee13-mod traditional} case. Here, in the time step immediately after the grid is isolated from the bulk electric system three areas are re-energized using DGs. However, in two of these areas, there is insufficient   DG  capacity to re-energize all loads so loads 675a and 701 are shed. In practice, such a grid state is impossible to operate unless technologies, such as smart meters, are available to disconnect these loads.

\subsection{Value of Modeling Load Shedding with Load Blocks}

Applying the \emph{block} DSR algorithm, we note a decrease in number of binary variables, and observe a substantial improvement (decrease) in solve time, but a notable increase in the number of loads and blocks shed over the restoration period (Table~\ref{results-table}). This is attributed to requiring that whole blocks be shed at a time, as opposed to the unrealistic assumption that individual loads can be shed within energized blocks. These results highlight the \emph{key contribution of this approach to modeling loads--1) modeling restoration without considering block level connectivity yields solutions that would be difficult (or impossible) to implement in practice and 2) modeling load restoration this way, while a more complex formulation, is more computationally efficient.} It is important to note that this approach to modeling load re-energizing does not preclude the traditional approach when the feeder has technologies to support individual load control. Switches placed at each load yields load blocks consisting of individual loads--indicating \emph{3) block is a generalization of traditional}. 

\subsection{Value of Modeling Grid-Forming Inverter Requirements}

The \emph{block} DSR algorithm still assumes that \emph{all}   DG  devices have grid-forming capability, which is typically not the case. Introducing the grid-forming inverter constraint (see \emph{block+gfm} cases in Table~\ref{results-table}), we note a significant increase in the number of binary variables, although still less than in the traditional case, and increases in the number of blocks (loads) shed, because some blocks that were previously energized in the \emph{block} case could not be energized with the additional requirement for grid-forming capabilities. While these results do tend to increase computation times, the results do show that the DSR can restore more load than is plausible when grid-forming inverters are not modeled.

\subsection{Scalability}

\begin{figure}
    \centering
    \includegraphics[width=0.6\columnwidth]{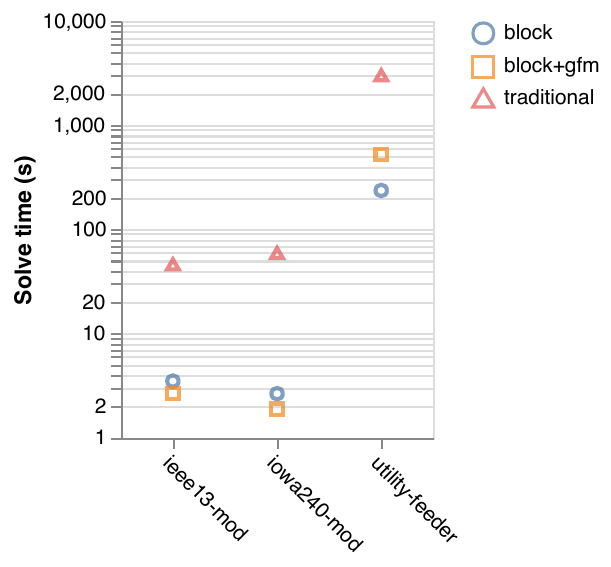}
    \caption{Solve times in seconds for the different cases and formulations.}
    \label{fig-solve-times}
\end{figure}

Finally, we further highlight that the \emph{block} DSR formulations provide significant benefits for problem scaling. 
As previously noted, for the larger \emph{utility-feeder} cases, the \emph{traditional} DSR formulation was not able to achieve similar optimality as compared to the other formulations at the time limit of $3000s$, whereas the \emph{block} DSR algorithms were able to find optimal solutions on the order of seconds to minutes. In Fig.~\ref{fig-solve-times}, the computational times are visualized on a log scale to better highlight how the traditional formulation takes at least an order of magnitude more time to find an optimal solution.

Interestingly, the \emph{block+gfm} is slightly computationally faster than \emph{block}--as reflected by the constraints reducing the space of feasible solutions.  However, this observation is reversed in the \emph{utility-feeder} case.  Given that \emph{block+gfm} introduces additional binary variables we would it expect it to be computationally less efficient than \emph{block}. We suspect that in the smaller cases, since the number of DGs is small and the introduction of additional binary is rather modest (2-3x more) and the time reduction due to a reduced space of feasible solutions dominates the time added by introducing binary variables.  This is contrast to \emph{utility-feeder}, which has many more DGs and introduces roughly 8x binary variables, driving an increase in solution time.

\section{Conclusion}\label{sec:conclusion}

This work introduces a new, more accurate, and better scaling formulation for the distribution system restoration problem that aggregates binary decision variables at the load block level, rather than assigning these variables to all objects within the blocks. We also introduce a new formulation for ensuring that energized blocks are supported by exactly one grid-forming inverter, based on a variant of graph coloring formulations. We find that the block formulation solves in at least an order of magnitude less time than traditional DSR, and produces more realistic and implementable results.

\section{Acknowledgments}

This work was performed with the support of the U.S. Department of Energy's (DOE) Office of Electricity's (OE) Microgrid Research and Development Program under the Resilient Operations of Networked Microgrids (RONM) and Dynamic Microgrids for Large-Scale DER Integration and Electrification (DynaGrid) projects and its OE program manager, Dan Ton. We gratefully acknowledge Dan's support of this work. The research work conducted at Los Alamos National Laboratory is done under the auspices of the National Nuclear Security Administration of the U.S. Department of Energy under Contract No. 89233218CNA000001.\\
\\
This manuscript is approved for unlimited public release (LA-UR-26-20496).

\appendix    
\section{Experiment Details}

The three use cases for the empirical results are: 
\begin{itemize}
    \item \textbf{Modified IEEE-13} (\emph{ieee13-mod}): This use case is a version of the IEEE-13 distribution feeder modified to add switches and DGs to form microgrids. It contains 25 buses (68 nodes), 7 switches, and 7 energy sources. The restoration period consists of 8 time steps. Details can be found in \cite{Fobes2020}, with an additional switch added at the substation since. 
    \item \textbf{Modified Iowa-240} (\emph{iowa240-mod}): This use case is a version of the Iowa-240 distribution feeder \cite{Iowa240} modified to add DGs to form microgrids. It contains 437 buses (770 nodes), 9 switches, and 8 energy sources, including the substation. The restoration period consists of 24 time steps. A more complete description can be found in \cite{Fobes2020}.
    \item \textbf{Utility Feeder} (\emph{utility-feeder}): This use case is a distribution feeder model
    provided by a utility partner. It contains 374 buses (998 nodes), 40 switches, and 26 energy sources consisting of a mixture of solar PV, battery energy storage devices, and a natural gas generator. The restoration period consists of 48 time steps. A more detailed description is found in \cite{Fobes2023}.
\end{itemize}

In all cases, the feeders are disconnected from the transmission system and must support themselves utilizing its DGs. For computational efficiency, empirical results were obtained by modeling constraint \ref{acopf:ohms_time} using the commonly used LinDistFlow approximation \cite{19266,7038399,7741261}. The results were generated using an x86\_64-pc-linux-gnu platform with 88 Logical Processors (44 Physical Cores) of Intel(R) Xeon(R) CPU E5-2699 v4 @ 2.20GHz with 252 GB RAM, using Julia v1.9.2 and Gurobi 10.0.2. Optimizations were terminated when an optimality gap of $0.0001$ was achieved, with a time limit of $3000$ seconds imposed.

In all cases, a limit of one switch closure per restoration time step was enforced, and loads (blocks) were required to remain energized for the remainder of the sequence once restored. In Table~\ref{results-table} we present the empirical results for the three use cases \emph{ieee13-mod}, \emph{iowa240-mod}, and \emph{utility-feeder} using formulations \emph{traditional}\footnote{The case where loads are fully energized or fully not energized.}, \emph{block}, and \emph{block} with grid-forming inverter constraints (\emph{block+gfm}).
For each case we list the number of binary variables, continuous variables, solve time, objective value, optimality gap, the total number of loads shed and the number of blocks shed over the whole restoration period, and the total active power delivered in MWh, denoted $P_d$.

\bibliographystyle{IEEEtran}
\bibliography{ref}

\end{document}